\documentclass[aps,prd,showpacs,preprint,nofootinbib,showkeys,preprintnumbers,floatfix]{revtex4}
\usepackage{graphicx,longtable}
\usepackage[dvips]{epsfig}
\usepackage{bm}

\begin{document}

\title{Unparticle Searches Through Gamma Gamma Scattering}

\author{O. \c{C}ak{\i}r}\email{ocakir@science.ankara.edu.tr}
\affiliation{Department of Physics, Ankara University, 06100 Tandogan,
Ankara, Turkey}
\author{K. O. Ozansoy}\email{oozansoy@physics.wisc.edu}
\affiliation{Department of Physics, University of Wisconsin, Madison,
WI 53706, USA}
\affiliation{Department of Physics, Ankara University, 06100 Tandogan,
Ankara, Turkey}

\date{\today}

\begin{abstract}
We investigate the effects of unparticles on
$\gamma \gamma\to \gamma \gamma$ scattering
for photon collider mode of the future
multi-TeV $e^+e^-$ linear collider.
We show the effects of unparticles on the
differential, and total scattering cross sections
for different polarization configurations.
Considering 1-loop Standard Model background contributions
from the charged fermions, and $W^{\pm}$ bosons to the
cross section, we calculate the upper limits on the unparticle couplings
$\lambda_{0}$ to the photons for various values of the
scaling dimension $d(1<d<2)$ at $\sqrt{s}=0.5-5$ TeV.
\end{abstract}

\medskip


\maketitle

\section{Introduction}

Recently, a mind-blowing, and very interesting,
new physics proposal has been presented
by Georgi \cite{Georgi:2007ek}.
According to this proposal, there could be a
scale invariant sector with a nontrivial infrared
fixed point living at a very high energy scale.
Since any theory with massive fields
cannot be scale invariant, the Standard Model(SM)
is not a scale invariant theory. Therefore,
such a scale invariant sector, if any,
should consist of massless fields and would interact with
the SM fields at the very high energies. One of the most
striking low energy properties of that proposal is that
using the low energy effective theory considerations one can
calculate the possible effects of such a scale invariant
sector for the TeV scale colliders.

In the Ref.\cite{Georgi:2007ek}, the fields of a very high
energy theory with a nontrivial fixed point are
called as BZ(for Banks-Zaks) fields according
to Ref.\cite{Banks:1981nn}.
Interactions of BZ operators ${\cal O}_{BZ}$ with
the SM operators ${\cal O}_{SM}$
are expressed by the exchange of
particles with a very high energy mass scale
${\cal M}_{\cal U}^k$ in the following form
\begin{eqnarray}
\label{1}
 \frac{1}{{\cal M}_{\cal U}^k}{O}_{BZ}{O}_{SM}
\end{eqnarray}
where BZ, and SM operators are defined as
${O}_{BZ}\in {\cal O}_{BZ}$ with mass dimension $d_{BZ}$,
and ${O}_{SM} \in {\cal O}_{SM} $ with mass dimension $d_{SM}$.
Low energy effects of the scale invariant ${\cal O}_{BZ}$ fields
imply a  dimensional transmutation. Thus, after
the dimensional transmutation Eq.(\ref{1}) is given as
\begin{eqnarray}
\label{2}
 \frac{C_{\cal U} \Lambda_{\cal U}^{d_{BZ}-d}}
{{\cal M}_{\cal U}^k}{O}_{\cal U}{O}_{SM}
\end{eqnarray}
where $d$ is the scaling mass dimension of the unparticle operator
$O_{\cal U}$ (in Ref.\cite{Georgi:2007ek}, $d=d_{\cal U}$ ),
and the constant $C_{\cal U}$ is a coefficient function.

Using the low energy effective field theory approach, very briefly
summarized above, in Refs \cite{Georgi:2007ek}, and \cite{Georgi:2007si}
main properties of the unparticle physics are presented.
A list of Feynman rules for the unparticles coupled to the SM
fields, and several implications of the collider phenomenology
are given in the Ref.\cite{Cheung:2007ue}. In this paper,
our calculations are based on the conventions of the
Ref.\cite{Cheung:2007ue}.

Searching for the new physics effects, the $e^+e^-$ linear
colliders have an exceptional advantageous for its appealing clean
background, and the possibility for the options of $e\gamma$, and
$\gamma\gamma$ colliders based on it.  Recently, for the new
physics searches, as a multi-TeV energy electron-positron
linear collider, the Compact Linear Collider(CLIC) proposed
and developed at CERN, is seriously taken into account. Numerous
works on the CLIC have been done so far \cite{clic}.
As other $e^+e^-$ linear colliders,
the CLIC would have the options for $e^-e^-,e\gamma$,
and $\gamma\gamma$ collider options, and possibilities of
polarized $e^+,e^-$ beams.
In this paper, we consider the $\gamma\gamma$ collider option
of the CLIC, to search for the unparticle physics effects.
Our results can easily be extended for other possible future
multi TeV-scale linear electron-positron colliders.
In Ref. \cite{Ginzburg:1983}, a detailed analysis on
$\gamma\gamma$ option of an $e^+e^-$ collider has been given.
Since $\gamma\gamma\to\gamma\gamma$ process can only occur
at loop-level in SM it gives a good opportunity to test of new
physics which has tree level contributions to the scattering
amplitude. Regarding this process, as new physics searches,
for example, supersymmetry \cite{gounaris1999},
large extra dimensions \cite{Cheung:1999ja,davoudiasl1999}, and
noncommutative space-time effects \cite{hewett2001}
has been taken into account. Here, we study the
effects of the unparticles on this process.

\section{Gamma Gamma Scattering}

The lowest order SM contributions to the
$\gamma\gamma\to\gamma\gamma$ process are
1-loop contributions of the charged fermions, and
$W^\pm$ bosons. In the limits,
for mandelstam parameters, $s,|t|,|u|>>M_W^2$,
and using certain symmetry arguments given
in the Ref.\cite{Jikia:1993tc,gounaris1999}
those 1-loop contributions can be expressed briefly.
We present the corresponding
1-loop SM amplitudes in the Appendix \ref{sec:a}.
Analysis of Fox {\it et al.} \cite{Fox:2007sy},
highlights that the
existence of the scalar unparticle
operator leads to the conformal
symmetry breaking when the Higgs
operator gets the vacuum expectation value.
If this symmetry breaking occurs at low
energies some strong constraints are
imposed on the unparticle sector.
Here, we assume that the effects of
unparticle sector on future high energy
collider energies could be measurable.

\begin{figure}
\includegraphics
{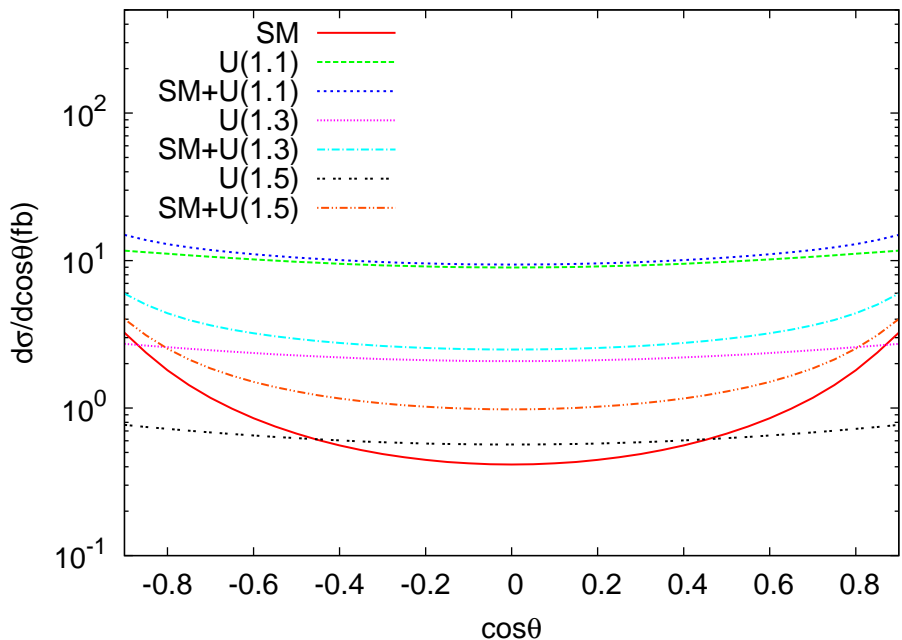} \caption{The unpolarized differential
cross sections with pure SM, and SM+${\cal U}$ effects at $\sqrt{s_{ee}}=1$ TeV. For the
unparticle effects,  $\lambda_0/\Lambda_{\cal U}=0.2$ TeV$^{-1}$,
$d=1.1$, $1.3$ and $d=1.5$.\label{fig:1}}
\end{figure}

\begin{figure}
\includegraphics
{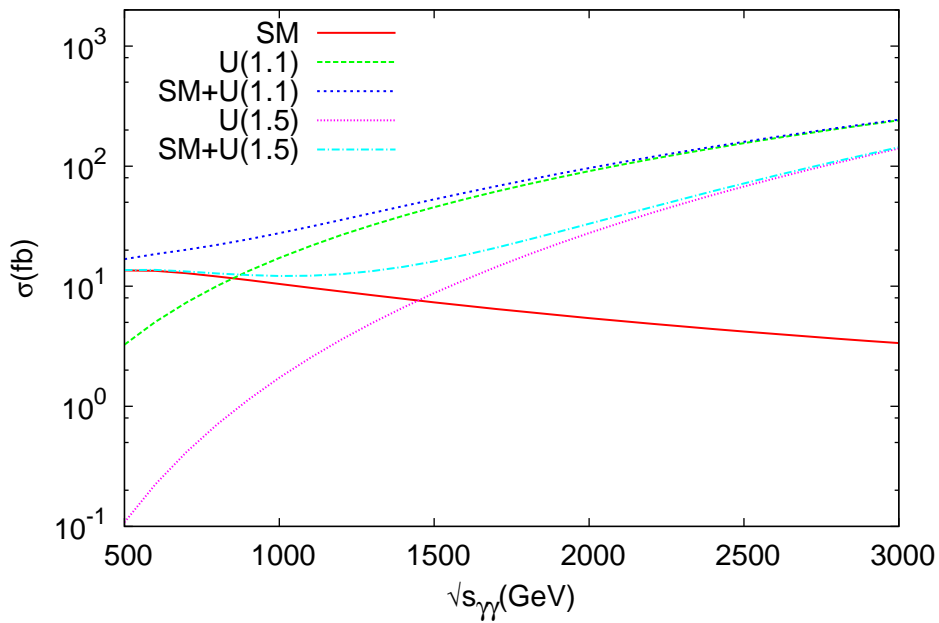} \caption{The unpolarized cross sections
for SM, and SM+${\cal U}$. For the unparticle
effects,  $\lambda_0/\Lambda_{\cal U}=0.1$ TeV$^{-1}$,
$d=1.1$, and $d=1.5$.\label{fig:2}}
\end{figure}

Using the low energy effective field theory assumptions
of Ref.s \cite{Georgi:2007si,Cheung:2007ue},
there are three tree level diagrams contributing to
$\gamma(p_1)\gamma(p_2)\to\gamma(p_3)\gamma(p_4)$
scattering amplitude from the exchange of the scalar
unparticle $U_S$ which can be expressed with the following amplitudes

\begin{eqnarray}
\label{3}
{M^s_{{\cal U}_S}} = && \Big[ \epsilon^{\mu }(p_1)\big[4i\frac{\lambda_0}{\Lambda_U^d}
[-p_1.p_2g_{\mu\nu}+p_{1\nu} p_{2\mu}]
\big]\epsilon^\nu(p_2)
\Big]
\Big[ \epsilon^{\rho *}(p_3)\big[4i\frac{\lambda_0}{\Lambda_U^d}
[-p_3.p_4g_{\rho\sigma}+p_{3\sigma} p_{4 \rho}]
\big]\epsilon^{\sigma *}(p_4)
\Big]
\nonumber\\
&&\times \big[\frac{i A_d}{2\sin{d\pi}}[-(p_1+p_2)^2]^{d-2} \big]
\end{eqnarray}

\begin{eqnarray}
\label{4}
{M^t_{{\cal U}_S}} =&& \Big[ \epsilon^{\rho *}(p_3)\big[4i\frac{\lambda_0}{\Lambda_U^d}
[p_1.p_3g_{\mu\rho}-p_{1\rho} p_{3\mu}]
\big]\epsilon^\mu(p_1)
\Big]
\Big[ \epsilon^{\sigma *}(p_4)\big[4i\frac{\lambda_0}{\Lambda_U^d}
[p_2.p_4g_{\nu\sigma}-p_{2\sigma} p_{4 \nu}]
\big]\epsilon^\nu(p_2)
\Big]
\nonumber\\
&&\times
\big[\frac{i A_d}{2\sin{d\pi}}[-(p_1-p_3)^2]^{d-2} \big]
\end{eqnarray}

\begin{eqnarray}
\label{5}
{M^u_{{\cal U}_S}}= &&
\Big[ \epsilon^{\sigma *}(p_4)\big[4i\frac{\lambda_0}{\Lambda_U^d}
[p_1.p_4g_{\mu\sigma}-p_{1\sigma} p_{4\mu}]
\big]\epsilon^\mu(p_1)
\Big]
\Big[ \epsilon^{\rho *}(p_3)\big[4i\frac{\lambda_0}{\Lambda_U^d}
[p_2.p_3g_{\nu\rho}-p_{2\rho} p_{3 \nu}]
\big]\epsilon^\nu(p_2)
\Big]
\nonumber\\
&&\times
\big[\frac{i A_d}{2\sin{d\pi}}[-(p_1-p_4)^2]^{d-2} \big]
\end{eqnarray}
with
\begin{equation}
\label{6}
A_d=\frac{16\pi^{5/2}}{{(2\pi)}^{2d}}
\frac{\Gamma(d+1/2)}{\Gamma(d-1)\Gamma(2d)}.
\end{equation}
where $\lambda_0$ and $\Lambda_U$ are the effective
coupling and the energy scale for scalar unparticle operator, respectively
\footnote{Very recently, after the first version of this paper
appeared online,
Grinstein {\it et al.} \cite{Grinstein:2008qk} have commented on several
issues related with the unparticle literature.
Besides the comments on
the scaling dimensions and the corrections in the form of the propagator
for vector and tensor unparticles, they have
pointed out that a generic unparticle scenario generates
contact interactions between particles.
Therefore, there could be generically a contribution
like, for example, our Eq. 5 but without a q-dependent propagator.
In our analysis we have not considered such contributions, in other words,
for very high energy physics effects due to the unparticle sector,
we consider only $O_{SM}O_{U}$ type interactions between unparticles and SM
particles, and not consider $O_{SM}O_{SM}$ type contact interactions between SM fields.}. We use the appropriate form of the scalar unparticle propagator,
$\Delta _F(q^2)=\frac{A_d}{2sin(d\pi)}(-q^2)^{d-2}$.
Since the mandelstam parameter $s>0$ there is a complex phase
factor due to s-channel amplitude. Thus, for s-channel propagator
one can consider $(-s^2)^{d-2}=(s)^{d-2}e^{-id\pi}$
\cite{Cheung:2007ue}. The implications of such a
complex factor could be studied only through the interference terms
\footnote{After we put the first
version of the present paper online, Chang {\it{et al.}} \cite{Chang:2008mk},
 have discussed the implications of this phase in the same context
of our paper.}.
The interesting features of this phase through s-channel
interferences between SM and unparticle amplitudes have been
discussed by \cite{Georgi:2007si}. In the calculations of the unpolarized and polarized
cross sections, we use the expressions given in the
Appendix. To give an idea about the unparticle effects on
the unpolarized differential cross section
$d\sigma/d\cos\theta$  with and without
unparticle effects is plotted in
Figure~\ref{fig:1}. In this figure, we choose
$\lambda_0/\Lambda_{\cal U}=0.2$ TeV$^{-1}$, and
the values $d=1.1$, $1.3$ and $d=1.5$ at $\sqrt{s_{ee}}=1$ TeV.
One can see from Fig.~\ref{fig:1},
the unparticle effect increases while the scaling dimension $d$
approaches to 1. The unpolarized total cross section
with respect to the center of mass energy of the
mono-energetic photon beams with and without
unparticle contributions is plotted in the Figure~\ref{fig:2}.
For the unparticle effects in that plot,
we assume, $\lambda_0/\Lambda_{\cal U}=0.1$ TeV$^{-1}$, and
we compare the shape of the distribution for $d=1.1$, and $d=1.5$.

For the polarized cross section calculations
of the back scattered photons, we
define $M_{ijkl}$ to be a helicity amplitude of
$\gamma\gamma\to\gamma\gamma$ scattering.
And, we use the following definitions
\begin{eqnarray}
\label{7}
|M(++)|^2 &=& \sum_{i,j}|M(++ij)|^2\\
|M(+-)|^2 &=& \sum_{i,j}|M(+-ij)|^2
\end{eqnarray}
where the summations are over the helicities of outgoing photons.
Therefore, depending on the initial fermion polarization
$P_e$, and the laser beam polarization $h_l$,
the differential scattering cross section in terms of the
average helicity $h_\gamma$ can be written as
\begin{eqnarray}
\label{8}
\frac{d\sigma}{d\cos\theta}&& =\frac{1}{(64\pi)}\int_{x_{1min}}^{0.83}dx_1
\int_{x_{2min}}^{0.83}dx_2
\frac{f(x_1)f(x_2)}{\hat s }
\nonumber\\
\times &&\Big[ \Big ( \frac{1+ h_{\gamma}(x_1)h_{\gamma}(x_2)}{2} \Big )
\Big |M^{SM+{\cal U}_S}(++) \Big |^2
+\Big ( \frac{1- h_{\gamma}(x_1)h_{\gamma}(x_2)}{2} \Big )
\Big |M^{SM+{\cal U}_S}(+-) \Big |^2\Big]
\end{eqnarray}
where $f(x)$ is the photon number density, and $h_{\gamma}$ is the average
helicity function presented in the Appendix \ref{sec:c},
and as $\sqrt{s_{ee}}\equiv\sqrt{s}$ being the center of mass energy
of the $e^{+}e^{-}$ collider, $\sqrt{\hat s}=\sqrt{x_1x_2 s_{ee}}$
is the reduced center of mass energy of the back-scattered photon beams,
and $x=E_\gamma/E_e$ is the energy fraction taken by the back-scattered photon beam.
In our analysis, we follow the usual collider assumptions
(for example Ref.s\cite{davoudiasl1999,hewett2001}) and
we take $|h_l|=1$, and $|P_e|=0.9$. Also, since we consider
the kinematical region $M_W^2/s,|M_W^2/t|,|M_W^2/u|<1$, in our analysis,
we use the cuts $\pi/6<\theta<5\pi/6$, and
$\sqrt{0.4}<x_i<x_{max}$ which have been used in the literature,
where $x_{max}$ is the maximum energy fraction of the back-scattered
photon, and its optimum value is 0.83.

\begin{figure}
\includegraphics
{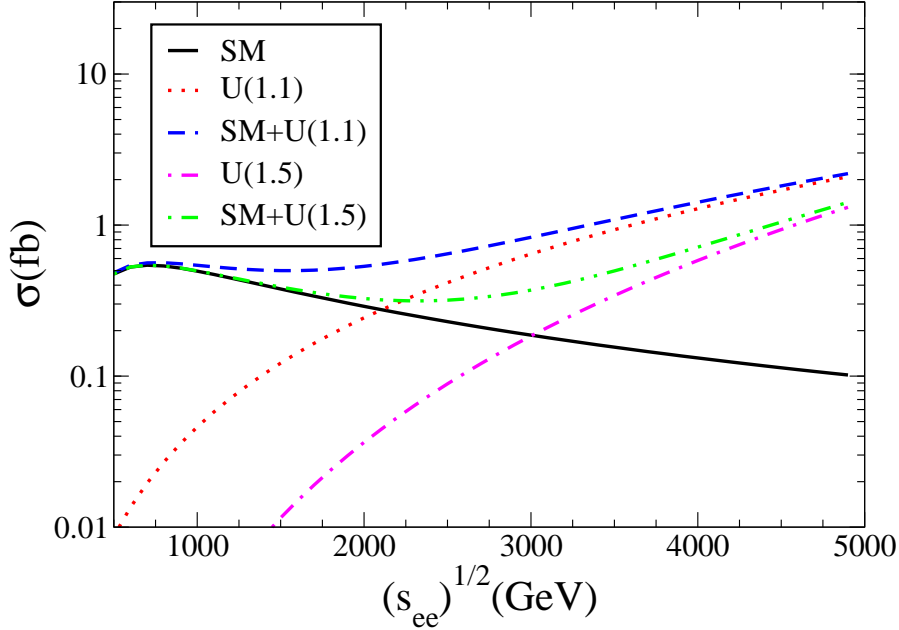} \caption{The
total polarized cross sections for SM, and SM+${\cal U}$ with the
polarization configuration $(++)$.
Here, we assume $\lambda_0/\Lambda_U=0.1$ TeV$^{-1}$, $d=1.1$ and $1.5$.\label{fig:3a} }
\end{figure}

\begin{figure}[tb]
\includegraphics
{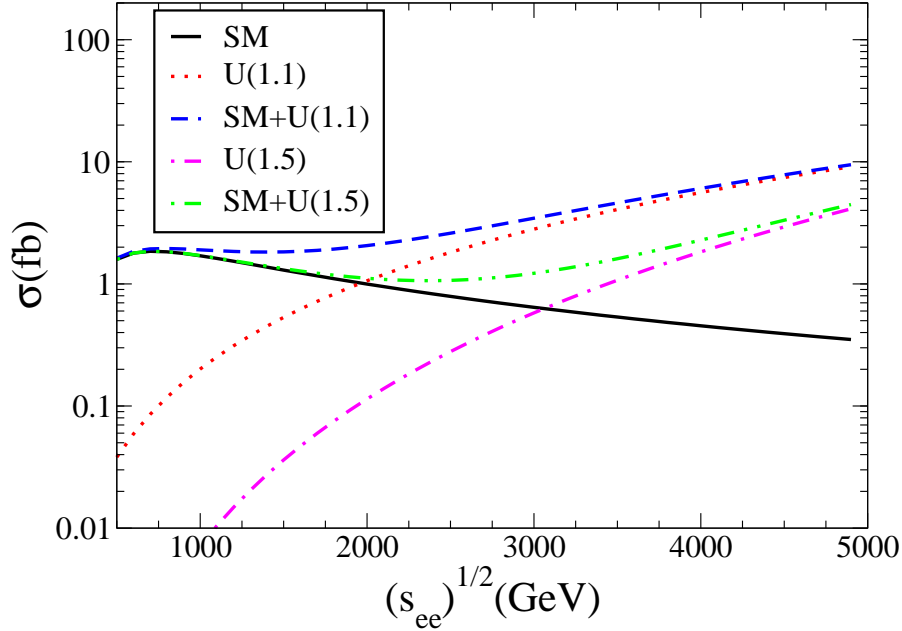} \caption{The total
polarized cross sections for SM, and SM+${\cal U}$
with the polarization configuration $(+-)$. Here, we
assume $\lambda_0/\Lambda_U=0.1$ TeV$^{-1}$, $d=1.1$ and $1.5$.\label{fig:3b}}
\end{figure}

In Figure \ref{fig:3a}, and Figure \ref{fig:3b},
to present schematic behavior of the polarized cross section
with or without unparticle contributions, we plot the total
cross section for two different polarization configurations
of initial beams. We use the following definitions
for the polarization configurations:
$(++)\equiv(++++)=(P_{e1}=0.9,h_{l1}=1;P_{e2}=0.9,h_{l2}=1)$,
and $(+-)\equiv(+-+-)=(P_{e1}=0.9,h_{l1}=-1;P_{e2}=0.9,h_{l2}=-1)$.
Those figures could give an idea about the scaling dimension $d$
dependence of the unparticle contribution.

\section{Limits}

Searching for the unparticle effects in a high
energy $\gamma\gamma\to\gamma\gamma$ scattering,
we extract the upper limits on the unparticle coupling
$\lambda_0$ regarding the $95\%$ C.L. analysis.
In the calculations, we use the standard
chi-square analysis for the following $\chi^2$ function

\begin{eqnarray}
\label{15}
 \chi^2=\sum_i\left[{{{d\sigma_i\over d\cos\theta}(SM)
-{d\sigma_i\over d\cos\theta}(SM+\cal{U})}
 \over\delta {d\sigma_i\over d\cos\theta}(SM)}\right]^2
\end{eqnarray}
\begin{figure}[tb]
\includegraphics{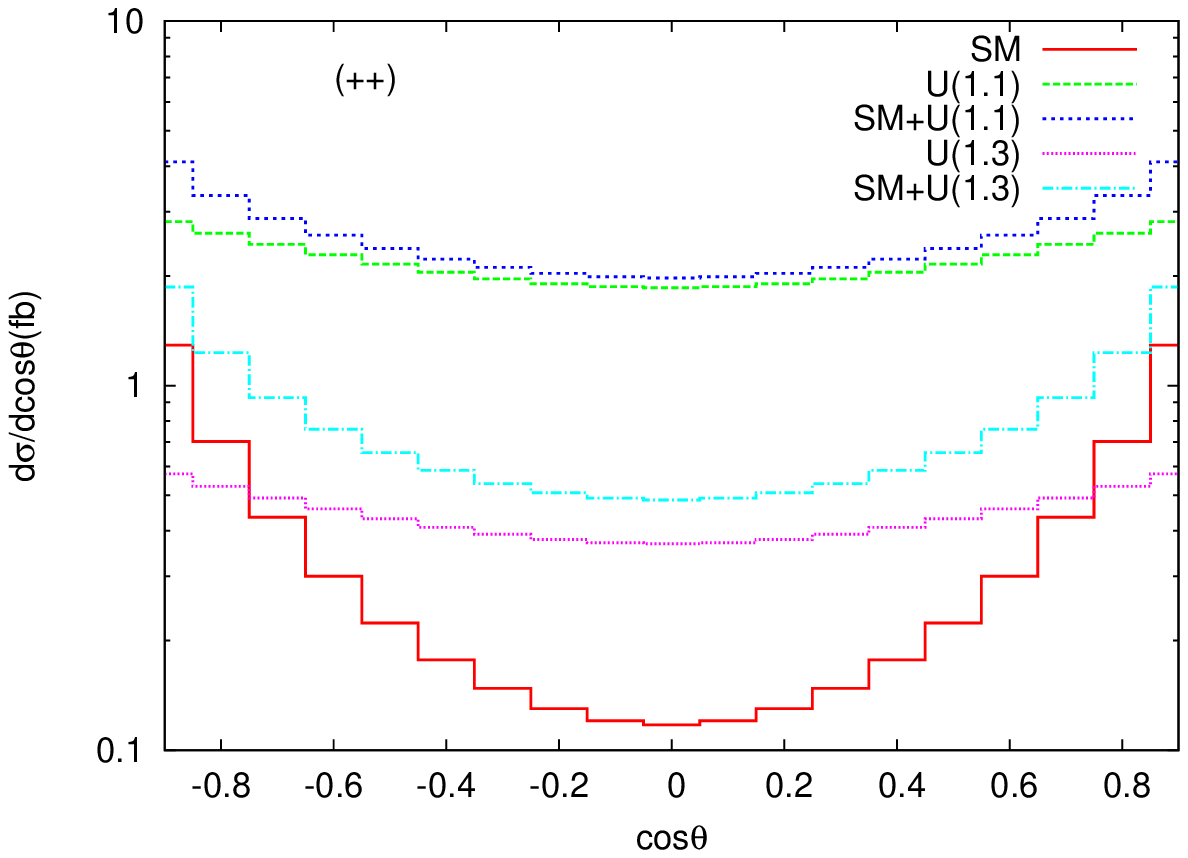}
\caption{The angular distribution for SM, and SM+${\cal U}$
cross sections with the polarization configuration $(++)$.
Here, we assume $\lambda_0/\Lambda_U=0.3$ TeV$^{-1}$,
$d=1.1$ and $d=1.3$ at $\sqrt{s_{ee}}=1$ TeV.\label{fig:4a}}
\end{figure}
\begin{figure}[tb]
\includegraphics{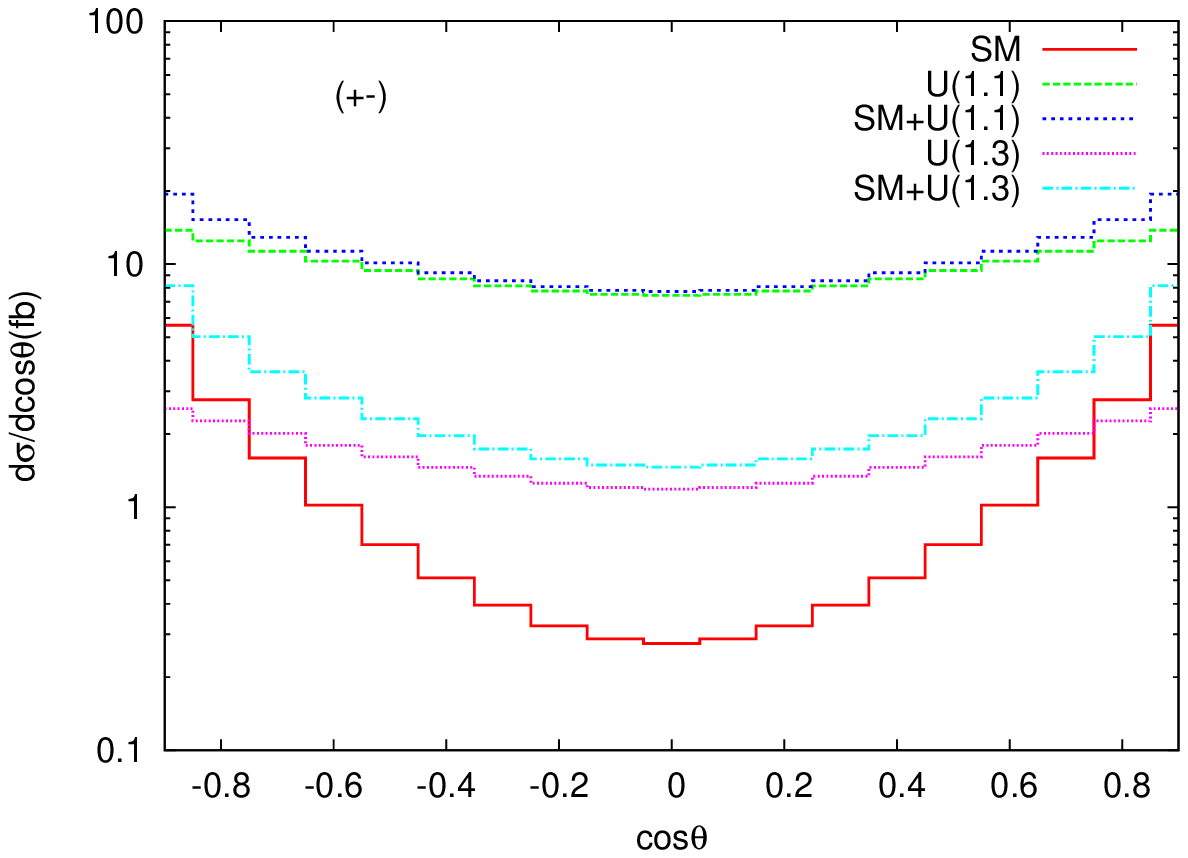}
\caption{The angular distribution for SM, and
SM+${\cal U}$ cross sections
with the polarization configuration $(+-)$. Here, we
assume $\lambda_0/\Lambda_U=0.3$ TeV$^{-1}$, $d=1.1$ and $d=1.3$ at $\sqrt{s_{ee}}=1$ TeV.\label{fig:4b}}
\end{figure}
where $\delta{d\sigma\over d\cos\theta}$ is the error on the measurement.
For one sided chi-square analysis, we assume $\chi^2\geq 2.7$, and
we take two possible luminosity values,
${\cal L}=100fb^{-1}$, and ${\cal L}=1000fb^{-1}$ per year.
We calculate the upper limits on the coupling of scalar unparticles by
performing a fit to binned photon angular distribution as shown
in Fig.~\ref{fig:4a} and \ref{fig:4b}.
For the signal and background calculation, we take into account only
the statistical error on the SM distribution. However,
the systematic errors should be considered
including $e^+/e^-$ beam conversion,
the photon-photon
collisions and the detector effects for the detection of photons,
if they are controlled well the limits can
be improved and benefitted from the advantage of high luminosity.
Our limits on $\lambda_0$ are presented in the
Table~\ref{tab1}, and Table~\ref{tab2} for two polarization configurations.

\begin{table}[tb]
{\caption{Upper limits on the $\lambda_0$ for the
polarization configuration(++) for  ${\cal L}=100(1000)$ fb$^{-1}$ and $\Lambda_U=1000$ GeV. \label{tab1}}}
\begin{ruledtabular}
\begin{tabular}{lcccccc}
$\sqrt{s}$ GeV &  d=1.01  & d=1.1 & d=1.3 & d=1.5 & d=1.7 & d=1.9 \\
\hline
500  &$0.0865(0.065)$ & $0.113(0.085)$ & $0.194(0.1455)$ & $0.316(0.237)$ & $0.479(0.359)$ & $0.551(0.413)$\\
1000 &$0.0605(0.0455)$ & $0.0745(0.056)$ & $0.1115(0.0835)$ & $0.1585(0.1185)$ & $0.2095(0.157)$ & $0.2105(0.1575)$\\
3000 &$0.0300(0.0225)$ & $0.0335(0.0250)$ & $0.0403(0.030)$ & $0.0458(0.0345)$ & $0.0485(0.0365)$ & $0.0393(0.0295)$\\
5000 &$0.0213(0.016)$ & $0.0225(0.017)$ & $0.0245(0.0185)$ & $0.0253(0.019)$ & $0.0243(0.0183)$ & $0.0178(0.0133)$
\end{tabular}
\end{ruledtabular}
\end{table}

\begin{table}[tb]
{\caption{Upper limits on the $\lambda_0$ for the
polarization configuration(+-)
for ${\cal L}=100(1000)$ fb$^{-1}$ and $\Lambda_U=1000$ GeV.\label{tab2}}}
\begin{ruledtabular}
\begin{tabular}{lcccccc}
$\sqrt{s}$ GeV &  d=1.01  & d=1.1 & d=1.3 & d=1.5 & d=1.7 & d=1.9\\
\hline
500  &$0.069(0.052)$ & $0.091(0.0685)$ & $0.1625(0.122)$ & $0.2765(0.2075)$ & $0.432(0.324)$ & $0.504(0.378)$\\
1000 &$0.0478(0.0358)$ & $0.0593(0.0443)$ & $0.0925(0.0695)$ & $0.1375(0.1033)$ & $0.1875(0.1408)$ & $0.1915(0.1435)$\\
3000 &$0.0235(0.0178)$ & $0.0265(0.0198)$ & $0.0333(0.0250)$ & $0.0398(0.0298)$ & $0.0435(0.0328)$ & $0.0358(0.0268)$\\
5000 &$0.0165(0.0125)$ & $0.018(0.0135)$ & $0.0205(0.0153)$ & $0.022(0.0165)$ & $0.0218(0.0163)$ & $0.0165(0.0123)$
\end{tabular}
\end{ruledtabular}
\end{table}

\begin{figure}[ht]
\includegraphics{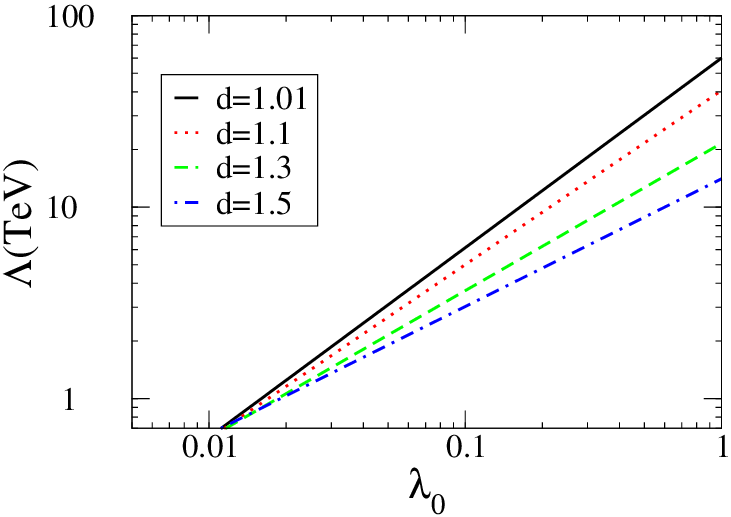}
\caption{Upper limits on the scalar unparticle
coupling $\lambda_0$ depending on $\Lambda_{\cal U}$ for (++)
polarization at CLIC 5 TeV energy.\label{fig:5a}}
\end{figure}

\begin{figure}[ht]
\includegraphics{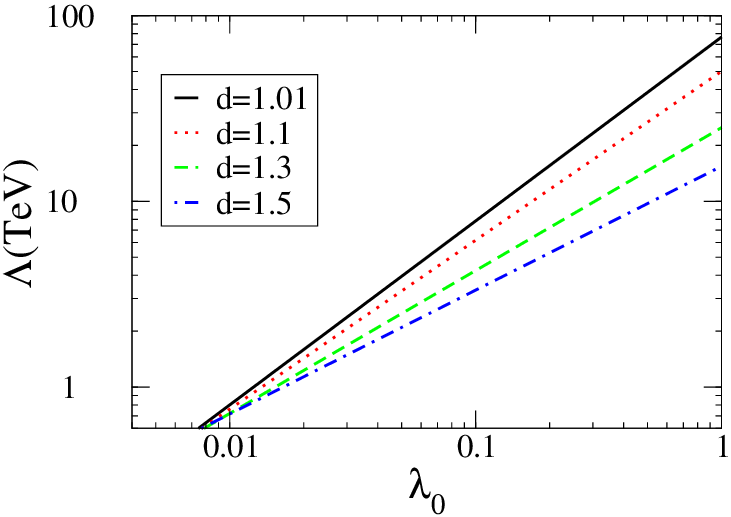}
\caption{Upper limits on the scalar unparticle
coupling $\lambda_0$ depending on $\Lambda_{\cal U}$ for (+-)
polarization at CLIC 5 TeV energy.\label{fig:5b}}
\end{figure}

Since the the cross section is proportional with
the $\lambda_0^4/\Lambda_{\cal U}^{4d}$, our
limits can be restated regarding the corresponding
behaviors of $\lambda_0$, and $\Lambda_{\cal U}$.
In Figure \ref{fig:5a} and \ref{fig:5b}, for the polarization
configuration $(++)$ and $(+-)$, we plot the corresponding
behaviors of $\Lambda_U$ and $\lambda_0$.
Right hand side of each curve is ruled out according
to the $95\%$C.L. analysis.
For the analysis schemes discussed above the similar results
can easily be obtained for the other center of mass energies with
low/high luminosities.

In conclusion, for different values of the scaling
dimension $d$, we put upper limits on $\lambda_0$ assuming
the scalar unparticle effects on the polarized cross section can
be distinguished from the SM contribution at $95\%$C.L. In our
analysis, we consider the multi-TeV CLIC electron-positron
collider, which will be launched at the CERN,
for the center of mass energies $\sqrt{s}=0.5$ TeV-$5.0$
TeV, and the luminosities ${\cal L}=100fb^{-1}$,
and ${\cal L}=1000fb^{-1}$ per year.
Our calculations show that the limits on $\lambda_0$
get more stringent as one increases the
luminosity and the center of mass energy of the collider.
Our limits are consistent with the limits calculated from other
low and high energy physics implications, for example
\cite{Anchordoqui:2007dp,Balantekin:2007eg}.

\section*{ACKNOWLEDGMENTS}
It is a pleasure to thank B. Balantekin for many helpful
conversations and discussions. KOO would like to thank
to the members of the Nuclear Theory Group of University of Wisconsin
for their hospitality, and acknowledges the
support through the Scientific and Technical Research
Council (TUBITAK) BIDEP-2219 grant. The
work of O. C. was supported in part by the State Planning
Organization (DPT) under grant no DPT-2006K-120470 and in part by
the Turkish Atomic Energy Authority (TAEA) under grant no
VII-B.04.DPT.1.05.

\appendix

\section{}

\subsection{\label{sec:a}1-loop SM Amplitudes}

The lowest order SM contributions to the
$\gamma\gamma\to\gamma\gamma$ process are
1-loop contributions of the charged fermions, and
$W^\pm$ bosons. There are 16  helicity amplitudes contributing
at the 1-loop level, and only three of them can be stated independently.
We can choose them $(++++),(+++-)$, and $(++--)$.
In the limits, $s,|t|,|u|>>M_W^2$,
the only significant contributions come from $(++++)$
polarization configuration, and that can be expressed
in the following form, Ref.\cite{Jikia:1993tc,gounaris1999}.
For the W boson contribution,
\begin{eqnarray}
\frac{M_{++++}^{(W)}(\hat{s},\hat{t},\hat{u})}{\alpha^{2}} & \approx & 12+12\left(\frac{\hat{u}-\hat{t}}{\hat{s}}\right)\left[\ln\left
(\frac{-\hat{u}-i\epsilon}{m_{W}^{2}}\right)-\ln\left
(\frac{-\hat{t}-i\epsilon}{m_{W}^{2}}\right)\right]\nonumber\\
 &  & +16\left(1-\frac{3\hat{t}\hat{u}}{4\hat{s}^{2}}\right)
\left\{ \left[\ln\left(\frac{-\hat{u}-i\epsilon}{m_{W}^{2}}\right)
-\ln\left(\frac{-\hat{t}-i\epsilon}{m_{W}^{2}}\right)
\right]^{2}+\pi^{2}\right\} \nonumber\\ &  & +16\hat{s}^{2}
\left[\frac{1}{\hat{s}\hat{t}}\ln\left(\frac{-\hat{s}-i\epsilon}{m_{W}^{2}}\right)
\ln\left(\frac{-\hat{t}-i\epsilon}{m_{W}^{2}}\right)
+\frac{1}{\hat{s}\hat{u}}\ln\left(\frac{-\hat{s}-i\epsilon}{m_{W}^{2}}\right)
\ln\left(\frac{-\hat{u}-i\epsilon}{m_{W}^{2}}\right)\right]\nonumber\\
 &  & +\frac{16\hat{s}^{2}}{\hat{t}\hat{u}}
\ln\left(\frac{-\hat{t}-i\epsilon}{m_{W}^{2}}\right)
\ln\left(\frac{-\hat{u}-i\epsilon}{m_{W}^{2}}\right)
\end{eqnarray}
for the fermion loop,
\begin{eqnarray}
\frac{M_{++++}^{(f)}(\hat{s},\hat{t},\hat{u})}{\alpha^{2}Q_{f}^{4}} & \approx & -8-8\left(\frac{\hat{u}-\hat{t}}{\hat{s}}\right)\left[\ln\left
(\frac{-\hat{u}-i\epsilon}{m_{f}^{2}}\right)-\ln\left
(\frac{-\hat{t}-i\epsilon}{m_{f}^{2}}\right)\right]\nonumber\\
&  & -4\left(\frac{\hat{t}^{2}+u^{2}}{s^{2}}\right)\left\{ \left[\ln\left(\frac{-\hat{u}-i\epsilon}{m_{f}^{2}}\right)
-\ln\left(\frac{-\hat{t}-i\epsilon}{m_{f}^{2}}
\right)\right]^{2}+\pi^{2}\right\}
\end{eqnarray}
where $Q_f$ is the fermion charge, $m_f$ is the mass of the fermion,
and  for the helicity amplitudes we use $M_{ h_{1} h_{2}h_{3}h_{4}}$
with the photon helicities $h_{i}=\pm$.
Using the assumptions given in \cite{gounaris1999},
the other significant helicity amplitudes can be
generated by using the relations
$M_{+-+-}(\hat{s},\hat{t},\hat{u})=M_{++++}(\hat{u},\hat{t},\hat{s})$
and $M_{+--+}(\hat{s},\hat{t},\hat{u})=M_{+-+-}(\hat{s},\hat{u},\hat{t})$.

\subsection{\label{sec:b}Expressions for unparticle contributions}

In the calculations, we assume the following center of mass
reference frame kinematical relations

\begin{eqnarray}
p_1^{\mu}=&&E(1,0,0,1),\quad p_2^{\mu}=E(1,0,0,-1)\\
p_3^{\mu}=&&E(1,\sin\theta,0,\cos\theta),\quad
p_4^{\mu}=E(1,-\sin\theta,0,-\cos\theta)\\
\epsilon_1^\mu=&&-\frac{1}{\sqrt{2}}(0,h_1,i,0)\quad
\epsilon_2^\mu=\frac{1}{\sqrt{2}}(0,-h_2,i,0)
\end{eqnarray}

where $\epsilon_1\equiv\epsilon_1(h_1),\epsilon_2\equiv\epsilon_1(h_2)$, etc.,
$h_1,h_2=\{ +,- \}$ stand for the polarizations, and we assume that the
summation is over the final state polarizations.

Therefore, one can find the following terms
\begin{eqnarray}
|M^s_{{\cal U}_S}(++)|^2=&&|M^s_{{\cal U}_S}(--)|^2=\frac{1}{8}[f(d)]^2[s]^{2d}\\
|M^s_{{\cal U}_S}(+-)|^2=&&|M^s_{{\cal U}_S}(-+)|^2=0\\
|M^t_{{\cal U}_S}(++)|^2=&&|M^t_{{\cal U}_S}(+-)|^2=|M^t_{{\cal U}_S}(-+)|^2=|M^t_{{\cal U}_S}(--)|^2=\frac{1}{16}[f(d)]^2[-t]^{2d}\\
|M^u_{{\cal U}_S}(++)|^2=&&|M^u_{{\cal U}_S}(+-)|^2=
|M^u_{{\cal U}_S}(-+)|^2=|M^u_{{\cal U}_S}(--)|^2=\frac{1}{16}[f(d)]^2[-u]^{2d}
\end{eqnarray}
The phase exp$(-id\pi)$ associates with the $s-t$ and $s-u$ channel interferences
\begin{eqnarray}
2{\cal R}e({M^s}^*_{{\cal U}_S}{M^t}_{{\cal U}_S})(++)=&&
2{\cal R}e({M^s}^*_{{\cal U}_S}{M^t}_{{\cal U}_S})(--)=
\frac{1}{8}[f(d)]^2[s]^{d}[-t]^{d} \cos (d\pi)\\
2{\cal R}e({M^s}^*_{{\cal U}_S}{M^t}_{{\cal U}_S})(+-)=&&
2{\cal R}e({M^s}^*_{{\cal U}_S}{M^t}_{{\cal U}_S})(-+)=0\\
2{\cal R}e({M^s}^*_{{\cal U}_S}{M^u}_{{\cal U}_S})(++)=&&
2{\cal R}e({M^s}^*_{{\cal U}_S}{M^u}_{{\cal U}_S})(--)=
\frac{1}{8}[f(d)]^2[s]^{d}[-u]^{d}\cos (d\pi)\\
2{\cal R}e({M^s}^*_{{\cal U}_S}{M^u}_{{\cal U}_S})(+-)=&&
2{\cal R}e({M^s}^*_{{\cal U}_S}{M^u}_{{\cal U}_S})(-+)=0\\
2{\cal R}e({M^t}^*_{{\cal U}_S}{M^u}_{{\cal U}_S})(++)=&&
2{\cal R}e({M^t}^*_{{\cal U}_S}{M^u}_{{\cal U}_S})(--)=
\frac{1}{8}[f(d)]^2[tu]^{d}\\
2{\cal R}e({M^t}^*_{{\cal U}_S}{M^u}_{{\cal U}_S})(+-)=&&
2{\cal R}e({M^t}^*_{{\cal U}_S}{M^u}_{{\cal U}_S})(-+)=0
\end{eqnarray}
where
\begin{equation}
\label{a1}
f(d)=\frac{8{\lambda_{0}^2} A_d}{ \Lambda^{2d}\sin (d\pi)},
\end{equation}

After the first version of this paper appeared online,
similar works have been appeared, \cite{Chang:2008mk, Kikuchi:2008pr}.
Our revised equations including the unparticle phase
are in agreement with those papers. If one takes average
over the squared helicity amplitudes  then gets
\begin{equation}
 |\bar M|^2=\frac{1}{4}[f(d)]^2
\big\{ [s]^{2d}+[-t]^{2d}+[-u]^{2d}
+[tu]^{d}+\big([s]^{d}[-t]^{d}+[s]^{d}[-u]^{d}\big)\cos(d\pi) \big\}.
\end{equation}

\subsection{\label{sec:c}Polarization Functions}

Let $h_e$ and $h_l$ be the polarizations of the electron beam and
the laser photon beam, respectively. According to \cite{Ginzburg:1983}, following function can be defined

\begin{eqnarray}
 C(x)=\frac{1}{1-x}+1-x-4r(1-r)-h_eh_lrz(2r-1)(2-x)
\end{eqnarray}
where $r=\frac{x}{z(1-x)}$, and $z=4E_e E_l /m_e^2$ describes the laser photon energy. Therefore, the photon number
density is given by

\begin{eqnarray}
 f(x,h_e,h_l,z)=\big(\frac{2\pi\alpha^2}{m_e^2z\sigma_c} \big)C(x)
\end{eqnarray}
where
\begin{eqnarray}
 \sigma_c=&&\big(\frac{2\pi\alpha^2}{m_e^2z} \big)
\big[\big(1-\frac{4}{z}-\frac{8}{z^2} \big)ln(z+1)
+\frac{1}{2}+\frac{8}{z}-+\frac{1}{2(z+1)^2} \big]\nonumber\\
&&+h_eh_l\big(\frac{2\pi\alpha^2}{m_e^2z} \big)
\big[\big(1+\frac{2}{z} \big)ln(z+1)-\frac{5}{2}+
\frac{1}{z+1}-\frac{1}{2(z+1)^2} \big]
\end{eqnarray}

The average helicity is given by

\begin{eqnarray}
h_\gamma(x,h_e,h_l,z)=\frac{1}{C(x)}
\big\{h_e\big[\frac{x}{1-x}+x(2r-1)^2 \big]
-h_l(2r-1)\big(1-x+\frac{1}{1-x} \big) \big\}
\end{eqnarray}


\begin{thebibliography}{99}

\bibitem{Georgi:2007ek}
  H.~Georgi,
  Phys.\ Rev.\ Lett.\  {\bf 98}, 221601 (2007)
  [arXiv:hep-ph/0703260].

\bibitem{Banks:1981nn}
  T.~Banks and A.~Zaks,
  Nucl.\ Phys.\  B {\bf 196}, 189 (1982).

\bibitem{Georgi:2007si}
  H.~Georgi,
  Phys.\ Lett.\  B {\bf 650}, 275 (2007)
  [arXiv:0704.2457 [hep-ph]].

\bibitem{Cheung:2007ue}
  K.~Cheung, W.~Y.~Keung and T.~C.~Yuan,
  Phys.\ Rev.\ Lett.\  {\bf 99}, 051803 (2007)
  arXiv:0704.2588 [hep-ph];
  Phys.\ Rev.\  D {\bf 76}, 055003 (2007)
  [arXiv:0706.3155 [hep-ph]].

\bibitem{Anchordoqui:2007dp}
  L.~Anchordoqui and H.~Goldberg,
  arXiv:0709.0678 [hep-ph].

\bibitem{Balantekin:2007eg}
  A.~B.~Balantekin and K.~O.~Ozansoy,
  Phys.\ Rev.\  D {\bf 76} (2007) 095014
  [arXiv:0710.0028 [hep-ph]].


\bibitem{Chang:2008mk}
  C.~F.~Chang, K.~Cheung and T.~C.~Yuan,
  arXiv:0801.2843 [hep-ph].

\bibitem{Kikuchi:2008pr}
  T.~Kikuchi, N.~Okada and M.~Takeuchi,
  arXiv:0801.0018 [hep-ph].


\bibitem{clic}

  E.~Accomando {\it et al.}  [CLIC Physics Working Group],
  arXiv:hep-ph/0412251.
  J.~A.~Aguilar-Saavedra {\it et al.}  [ECFA/DESY LC Physics Working Group],
  arXiv:hep-ph/0106315.
  R.~W.~Assmann {\it et al.}, {\it A 3-TeV e+ e- linear
collider based on CLIC technology},
  CERN-2000-008, Geneva, 2000.
  R.~W.~Assmann {\it et al.}, {\it CLIC contribution to the
  technical review committee on a 500 GeV $e^+ e^-$
  linear collider}, CERN-2003-007, Geneva, 2003.
  A.~De Roeck,
  arXiv:hep-ph/0311138.

\bibitem{Ginzburg:1983}
  I.~F.~Ginzburg, G.~L.~Kotkin, V.~G.~Serbo and V.~I.~Telnov,
  Nucl.\ Instrum.\ Meth.\  {\bf 205} (1983) 47.
  I.~F.~Ginzburg, G.~L.~Kotkin, S.~L.~Panfil, V.~G.~Serbo and V.~I.~Telnov,
  Nucl.\ Instrum.\ Meth.\  A {\bf 219}, 5 (1984).

\bibitem{Dawson:2004xz}
  S.~Dawson and M.~Oreglia,
  Ann.\ Rev.\ Nucl.\ Part.\ Sci.\  {\bf 54}, 269 (2004)
  [arXiv:hep-ph/0403015].


\bibitem{Jikia:1993tc}
  G.~Jikia and A.~Tkabladze,
  Phys.\ Lett.\  B {\bf 323} (1994) 453
  [arXiv:hep-ph/9312228].

\bibitem{gounaris1999}
  G.~J.~Gounaris, P.~I.~Porfyriadis and F.~M.~Renard,
  Eur.\ Phys.\ J.\  C {\bf 9} (1999) 673
  [arXiv:hep-ph/9902230].
;Phys. Lett. B, 452, 76(1999)

\bibitem{Cheung:1999ja}
  K.~m.~Cheung,
  Phys.\ Rev.\  D {\bf 61} (2000) 015005
  [arXiv:hep-ph/9904266].

\bibitem{davoudiasl1999}
  H.~Davoudiasl,
  Phys.\ Rev.\  D {\bf 60}, 084022 (1999)
  [arXiv:hep-ph/9904425].

\bibitem{hewett2001}
  J.~L.~Hewett, F.~J.~Petriello and T.~G.~Rizzo,
  Phys.\ Rev.\  D {\bf 64} (2001) 075012
  [arXiv:hep-ph/0010354].

\bibitem{Fox:2007sy}
  P.~J.~Fox, A.~Rajaraman and Y.~Shirman,
  Phys.\ Rev.\  D {\bf 76} (2007) 075004
  [arXiv:0705.3092 [hep-ph]].

\bibitem{Grinstein:2008qk}
  B.~Grinstein, K.~Intriligator and I.~Z.~Rothstein,
  arXiv:0801.1140 [hep-ph].


\bibitem{Yao:2006px}
  W.~M.~Yao {\it et al.}  [Particle Data Group],
  J.\ Phys.\ G {\bf 33}, 1 (2006).

\end{thebibliography}
\end{document}